\documentclass[twocolumn]{aastex631}

\usepackage{xcolor}
\usepackage{graphicx}%
\usepackage{amsmath}%
\usepackage{dcolumn}%
\usepackage{bm}%
\usepackage{times}%
\usepackage{hyperref}%
\usepackage[normalem]{ulem}
\hypersetup{
  colorlinks=true,        %
  linkcolor=blue,         %
  citecolor=cyan,         %
  urlcolor=cyan            %
}

\newcommand{\nsat}{\ensuremath{n_\text{sat}}}

\newcommand{\ep}{\varepsilon}

\shorttitle{Bulk viscosity in neutron star mergers}
\shortauthors{E.R. Most et al.}

\begin{document}

\title{Emergence of microphysical bulk viscosity in binary neutron star post-merger dynamics}

\author[0000-0002-0491-1210]{Elias R. Most}
\affiliation{TAPIR, Mailcode 350-17, California Institute of Technology, Pasadena, CA 91125, USA}
\affiliation{Walter Burke Institute for Theoretical Physics, California Institute of Technology, Pasadena, CA 91125, USA}
\author{Alexander Haber}
\affiliation{Physics Department, Washington University in Saint Louis, 63130 Saint Louis, MO, USA}
\author{Steven P. Harris}
\affiliation{Institute for Nuclear Theory, University of Washington, Seattle, WA 98195, USA}
\author{Ziyuan Zhang}
\affiliation{Physics Department, Washington University in Saint Louis, 63130 Saint Louis, MO, USA}
\affiliation{McDonnell Center for the Space Sciences, Washington University in St. Louis, St. Louis, MO 63130, USA}
\author{Mark G. Alford}
\affiliation{Physics Department, Washington University in Saint Louis, 63130 Saint Louis, MO, USA}
\author{Jorge Noronha}
\affiliation{Illinois Center for Advanced Studies of the Universe \& Department of Physics, 
University of Illinois at Urbana-Champaign, Urbana, IL, 61801, USA}

\begin{abstract}
In nuclear matter in isolated neutron stars, the flavor content (e.g.,
proton fraction) is subject to weak interactions, establishing flavor
($\beta$-)equilibrium. However, there can be deviations from this
equilibrium during the merger of two neutron stars.  We study the resulting
out-of-equilibrium dynamics during the collision by incorporating direct
and modified Urca processes (in the neutrino-transparent regime) into general-relativistic hydrodynamics
simulations  with a simplified neutrino transport scheme.
We demonstrate how weak-interaction-driven bulk viscosity in post-merger simulations can emerge and assess the bulk viscous dynamics of the resulting flow. We further place limits on the impact of the post-merger gravitational wave strain. Our results show that weak-interaction-driven bulk viscosity can potentially lead to a phase shift of the post-merger gravitational wave spectrum, although the effect is currently on the same level as the numerical errors of our simulation.
\end{abstract}


\section{Introduction}
Binary neutron star mergers
\citep{LIGOScientific:2017vwq,LIGOScientific:2020aai} offer exciting
prospects for constraining the dense matter equation of state (EoS)
\citep{Lattimer:2015nhk,Oertel:2016bki,Ozel:2016oaf} (e.g.,
\citet{Flanagan:2007ix,Read:2009yp,Raithel:2019uzi,HernandezVivanco:2019vvk,Landry:2020vaw,Chatziioannou:2021tdi}).
This can be done either by using the tidal deformability encoded in the
inspiral gravitational wave signal
\citep{Bauswein:2017vtn,Annala:2017llu,Most:2018hfd, LIGOScientific:2018cki,
Raithel:2018ncd, De:2018uhw, Chatziioannou:2018vzf, Carson:2018xri}, the
potential presence of very massive neutron stars in gravitational wave
events
\citep{Fattoyev:2020cws,Most:2020bba,Tan:2020ics,Dexheimer:2020rlp,Tews:2020ylw,Nathanail:2021tay,Tan:2021ahl},
or by using the combined information from electromagnetic counterparts
\citep{Radice:2017lry,Margalit:2017dij, Rezzolla:2017aly, Ruiz:2017due,
Shibata:2019ctb}.  Critically, some of these constraints heavily rely on
properties of the mass ejected after the merger (e.g.,
\citealt{Cowperthwaite:2017dyu,Chornock:2017sdf,Villar:2017wcc,Nicholl:2017ahq,Troja:2018ruz,Tanvir:2017pws,Drout:2017ijr}),
which crucially depends on models for the post-merger evolution, including
the lifetime of the remnant neutron star (see, e.g.,
\citealt{Baiotti:2016qnr,Metzger:2019zeh,Radice:2020ddv} for recent reviews).
Future detection of post-merger gravitational wave signals
\citep{Shibata:2005xz,Punturo:2010zz,Reitze:2019iox,Ackley:2020atn,Wijngaarden:2022sah,Breschi:2022ens,Yu:2022upc}
have the potential to put even more stringent constraints on the EoS
\citep{Bauswein:2012ya,Bauswein:2014qla,Bernuzzi:2015rla,Takami:2014zpa,Takami:2014tva,Rezzolla:2016nxn,Bose:2017jvk,Vretinaris:2019spn,Raithel:2022orm,Breschi:2022xnc}.

There are various uncertainties in merger simulations, such as the precise form of the cold EoS, finite-temperature effects
\citep{Bauswein:2010dn,Figura:2020fkj,Raithel:2021hye}, and the appearance of
exotic degrees of freedom or phase transitions
\citep{Oechslin:2004yj,Most:2018eaw,Bauswein:2018bma,Most:2019onn,Weih:2019xvw,Chatziioannou:2019yko,Prakash:2021wpz,Tootle:2022pvd,Tan:2021ahl}.
In this work, we focus on a dynamical effect,
the equilibration of the proton fraction via weak interactions \citep{Alford:2017rxf,Radice:2021jtw}, and its potential to alter the post-merger dynamics and systematically bias our ability to infer dense matter properties \citep{Most:2021zvc}. 
This is possible because the dynamical timescales of a merger can be fast enough to drive the system out of $\beta$-equilibrium \citep{Hammond:2021vtv,Most:2021ktk}.
One expected
consequence of the resultant
re-equilibration via weak interactions 
is bulk viscosity \citep{Alford:2019qtm,Gavassino:2020kwo}.

Previous work has largely been devoted to understanding the properties
\citep{Alford:2019qtm,Alford:2019kdw} and the potential appearance of bulk
viscosity \citep{Alford:2017rxf,Most:2021zvc,Hammond:2021vtv} in the
remnant, or to study the idealized limit of perfect $\beta-$equilibrium
\citep{Ardevol-Pulpillo:2018btx,Hammond:2022uua}. While effects of
microphysical viscosity in neutron-star merger simulations have largely
been ignored, they have been well-studied in the related context of dense
matter formed heavy-ion collisions
\citep{Monnai:2009ad,Song:2009rh,Bozek:2009dw,Dusling:2011fd,Noronha-Hostler:2013gga,
Ryu:2015vwa,Ryu:2017qzn}, see also \citet{Romatschke:2017ejr} for a  review.

Going beyond all of the above, in this paper, we take essential steps
towards 
fully self-consistent numerical simulations of the out-of-equilibrium
dynamics of the flavor content of the matter in
the post-merger phase of a neutron star collision. We begin by briefly
reviewing the relevant processes (Urca processes) active in the core of
neutron stars \citep{1995A&A...297..717Y}. We then provide a detailed
discussion of the dynamics of relaxation to  $\beta-$equilibrium after the
merger. For the first time, we directly demonstrate the {emergence of
   bulk viscosity \citep{Gavassino:2023xkt}} in post-merger simulations
  and provide a first quantification of its effects. We conclude by
  discussing how this may alter the post-merger gravitational wave
  emission.\\

\section{Methods}
  We model the global dynamics of the system using general-relativistic
  (magneto-)hydrodynamics \citep{Duez:2005sf} in a dynamically evolved
  spacetime in the Z4c formulation \citep{Hilditch:2012fp}.  
  We evolve the baryon current $J_B^\mu = n_B u^\mu$ and the
  perfect-fluid energy-momentum tensor $T^{\mu\nu}$
  \citep{denicol2021microscopic} according to
\begin{align}
   \nabla_\alpha J_B^\alpha=0\,, ~~~ \nabla_\alpha T^{\alpha\beta} = Q_{\nu}^\beta \,,
   \label{eqn:hydro}
\end{align}
where  $Q_{\nu}^\beta$ is an effective energy ``sink" term associated with neutrino energy-losses \citep{Sekiguchi:2012uc,Galeazzi:2013mia}. 
Here, $n_B$ and $u^\mu$ are the baryon number density and fluid four-velocity, respectively. 

Assuming $npe$ matter, the electron fraction, $Y_e = n_e/n_B$, can only change by
weak-interaction decays of neutrons $(n)$ and protons $(p)$. We incorporate
these as follows.
First, we include a standard set of weak interactions without Urca contributions,
  $\mathcal{R}_{\nu_e}$, commonly used in leakage scheme approaches to
  neutron star mergers \citep{1996A&A...311..532R,2003MNRAS.342..673R}. 
  These include approximate direct Urca reactions 
  $ n + \nu_e \leftrightarrow p + e^- $ (though not $ n \to p + e^- + \bar{\nu}_e$ which is relevant at low neutrino density)
  which we replace as outlined in the following), pair annihilation $(e^+ + e^- \rightarrow \bar{\nu_e} + {\nu_e})$, plasmon decay $\gamma \rightarrow \bar{\nu}_e + \nu_e$. Those rates are taken from \citet{1996A&A...311..532R} and will, at low neutrino densities, not drive the system to flavor equilibrium.
    Secondly, we consistently include direct, ${\Gamma}_\text{dUrca}$, and
    ---for the first time--- modified, ${\Gamma}_\text{mUrca}$, Urca (net) rates \citep{Yakovlev:2000jp,Alford:2021ogv} that
  operate in the dense nuclear matter core of the merger; see our discussion of the neutrino transparent regime below. To track the relative number of protons, we evolve the relative fraction of electrons $(e)$
\begin{align}
   n_B \,u^\mu \nabla_\mu Y_e = \Gamma_\nu \equiv \mathcal{R}_{\nu_e} +{\Gamma}_{\rm mUrca} + {\Gamma}_{\rm dUrca}. 
    \label{eqn:Ye_evol}
\end{align}
We solve the coupled system of equations using the \texttt{Frankfurt/IllinoisGRMHD} code \citep{Most:2019kfe,Etienne:2015cea}, which utilizes the \texttt{EinsteinToolkit} infrastructure \citep{Loffler:2011ay}, where the Urca processes are included via a fully implicit operator-split approach.
The \texttt{FIL} code solves the equations of general-relativistic magnetohydrodynamics in dynamical spacetimes \cite{Duez:2005sf}. Since magnetic fields do not affect chemical equilibration, we set the magnetic field to zero throughout the evolution, making our simulations fully hydrodynamic. The system is integrated numerically using a fourth-order accurate version of the ECHO scheme \citep{DelZanna:2007pk}. The spacetime is evolved in the Z4c formulation of Einstein's equations \citep{Bernuzzi:2009ex,Hilditch:2012fp}, using a fourth-order accurate finite-difference discretization \citep{Zlochower:2005bj}.
We solve the equation on a set of eight nested grids, with a finest resolution of $262\, \rm m$ and an outer domain size of $3022\, \rm km$.
In addition, we also perform simulations at $150\,\rm m$ resolution when quantifying the numerical error of our results.
More details on the numerical setup and the choice of initial data can be found in \citep{Most:2021ktk}.

This first study considers a system with fixed binary parameters loosely modeled after the GW170817 event \citep{LIGOScientific:2017vwq}. We fix the total mass $M=2.74\, M_\odot$ with a mass ratio of $q=0.85$. This results in a hot, (meta-)stable neutron star remnant after the merger, which does not immediately collapse to a black hole \citep{Bauswein:2013jpa,Koppel:2019pys,Bauswein:2020aag,Tootle:2021umi,Kashyap:2021wzs,Kolsch:2021lub,Perego:2021mkd}. A detailed description of the (post-)merger dynamics and gravitational wave emission of this system for different EoSs can be found in \citep{Most:2021ktk,Raithel:2022orm}. This work focuses on the impact of departures from and relaxation back to beta equilibrium
\citep{Hammond:2021vtv,Most:2021ktk}. 

To approximate the complex dynamics of neutrino populations, one can divide the merger into two
regions: the neutrino-trapped regime and the neutrino-transparent regime. In the former, local opacities are so high that the 
neutrinos cannot escape, forming a gas. This leads to a correction
in the fluid element's overall pressure, energy density, and lepton number. 
Our work mainly focuses on the merger in the neutrino transparent regime,
but in the Appendix we provide an assessment of neutrino trapping
which will establish chemical equilibrium much faster than in the free
streaming regime \citep{Alford:2019kdw,Alford:2021lpp,Perego:2019adq}.
Accordingly, we treat neutrinos numerically in a simplified Leakage manner \citep{Galeazzi:2013mia}, assuming the free-streaming loss of neutrinos from each fluid element.
While this will not affect most of our qualitative conclusions on how
microphysical dissipation emerges in the post-merger dynamics, it will
change their quantitative impact, especially in macroscopic observables
such as the gravitational wave emission. We estimate the effect of neutrino transparency using additional simulations in the fully neutrino trapped regime--- see the Appendix for details.

In the neutrino transparent regime
beta equilibration occurs via the
direct Urca (dUrca), $n \rightarrow p + e^- + \bar{\nu}_e\,, ~~~ p+ e^- \rightarrow n + {\nu}_e\,$, and modified Urca (mUrca) processes, $n + X \rightarrow p + X + e^- + \bar{\nu}_e\,, ~~~ p+ e^- +X  \rightarrow n + X+ {\nu}_e\,$, where $X$ is a spectator nucleon \citep{1995A&A...297..717Y,Yakovlev:2000jp}. 
 For this first investigation of $\beta$-equilibration in merger dynamics, we use simple analytic expressions for the Urca rates obtained via the Fermi surface approximation in which all participating nucleons and electrons are assumed to be close to their Fermi surfaces. 
 Realistically, there are finite temperature corrections
 that are neglected here because
 their main effect is to
shift the temperature at which the resonant bulk viscosity reaches its maximum value
by a few MeV \citep{Alford:2018lhf,Alford:2021ogv}.

The neutrino-transparent regime of $npe$ matter can then be divided
into two sub-regimes, according to whether the
proton fraction $Y_e$ is above or below the
direct Urca threshold value of $0.11$ \citep{Yakovlev:2000jp,Lattimer:1991ib}.
Above this threshold, the faster direct Urca process dominates over modified Urca. Below the threshold, direct Urca is suppressed
and beta equilibrium occurs via the slower modified Urca process.
To probe the influence of the dUrca threshold,
we consider two EoSs, TMA \citep{Hempel:2009mc} and SFHo \citep{Steiner:2012rk,Alford:2021ogv}. At $\beta$-equilibrium, TMA has a direct Urca threshold just below $2\nsat$ (where $\nsat\approx 0.16 \, \rm fm^{-3}$ is nuclear saturation density) while in SFHo the proton fraction never reaches the direct Urca threshold
(in both cases referring to $\beta$-equilibrated matter at $T=0$).
More information can be found in the Appendix.

In the Fermi surface approximation, one can characterize the degree to which neutrino-transparent material is out of beta equilibrium by the chemical potential imbalance $\delta \mu = \mu_n - \mu_p -\mu_e$, where $\mu_a$ is the chemical potential
for particle $a$. 
In beta equilibrium $\delta\mu=0$ and
$Y_e=Y^\text{eq}(n,\ep)$, where $\ep$ is the total fluid energy density. Given that $\delta \mu = \delta \mu(\varepsilon, n_B,Y_e)$, one can use \eqref{eqn:hydro} and \eqref{eqn:Ye_evol} to find that $\delta \mu$ evolves according to the following equation
\begin{align}
   \tau\,u^\nu \nabla_\nu \delta \mu  +\delta \mu= - \,\mathcal{B}\, \tau\, \theta  + \mathcal{S}
    \label{eqn:PpPinew}
\end{align}
where $\tau = \Bigl(-\frac{\partial \ln\delta \mu}{\partial Y_e}\big|_{\varepsilon, n_B} \frac{\Gamma_\nu}{n_B}\Bigr)^{-1}$ is the flavor relaxation timescale, $\theta = \nabla_\mu u^\mu$ is the fluid expansion rate, $\mathcal{B} = \left.n_B\frac{\partial \delta \mu}{\partial n_B}\right|_{\varepsilon, Y_e} + \left.\left(\varepsilon + P\right)  \frac{\partial \delta \mu}{\partial \varepsilon}\right|_{n_B, Y_e}$ is a thermodynamic quantity, with $P = P(\varepsilon, n_B,Y_e)$ being the total pressure. Also, $\mathcal{S}= - \tau\,u_\nu Q^\nu\frac{\partial \delta \mu}{\partial \varepsilon}\Bigl|_{n_B, Y_e}$ denotes a source term coming from neutrino radiative energy loss. Equation \eqref{eqn:PpPinew} is a complicated, highly nonlinear\footnote{Note that $\tau,\mathcal{B},\mathcal{S}$ depend on $\delta\mu$ via the EoS and the reaction rates.} relaxation-type equation that describes how $\delta \mu$ varies given the local fluid expansion rate and the source. We shall show in the following that during a neutron star merger, the full relaxation equation \eqref{eqn:PpPinew} experiences different regimes that are nothing but different manifestations of the physics of bulk viscosity.

For slight deviations from beta equilibrium, $\Gamma_\nu \sim -\lambda_0 \delta \mu$, $1/\tau\to 1/\tau_0 = \frac{\partial \delta \mu}{\partial Y_e}\Bigr|_{\varepsilon, n_B,Y^\textrm{eq}} \frac{\lambda_0}{n_B}$, $\mathcal{B},\mathcal{S}\to \mathcal{B}_0,\mathcal{S}_0$ (at $\delta \mu=0$). Eq.~\eqref{eqn:PpPinew} 
enters the linear response regime where $\delta\mu$ relaxes to zero on a timescale $\tau_0$,
and the
relaxation dynamics can be understood in terms of bulk viscosity \citep{Gavassino:2020kwo,Camelio:2022ljs,Celora:2022nbp,Most:2021zvc}. Indeed, one can define the bulk scalar pressure correction, $\Pi$, by splitting the total pressure, $P$, into
\begin{align}
   P(n_B,\ep,Y_e) = P^{\rm eq} (n_B,\ep) + \Pi(n_B,\ep,Y_e) \,,
   \label{eqn:PpPi}
\end{align}
where $P^{\rm eq} = P(n_B,\ep, Y_e=Y_e^{\rm eq})$ is the pressure in beta equilibrium. In the linear response regime, one can approximate $\Pi \sim I_1\delta \mu$ (with $I_1 = \left.\frac{\partial P}{\partial \delta \mu}\right|_{\varepsilon, n_B, \delta \mu = 0}$) to find \citep{Gavassino:2020kwo} 
\begin{align}
   \tau_0 u^\mu \nabla_\mu \Pi + \Pi= -\zeta_0 \theta +\mathcal{S}_0^\textrm{bulk},
    \label{eqn:PpPibulk}
\end{align}
where $\zeta_0 =  \mathcal{B}_0 I_1 \tau_0$ is the static bulk viscosity coefficient induced by the weak-interactions \citep{Sawyer:1989dp,Sad:2009hba,Harris:2020rus}. For periodic density oscillations with frequency $\omega$, $\delta \mu$ and $\Pi$ also oscillate and \eqref{eqn:PpPibulk} defines an %
AC bulk viscosity $\zeta(\omega) = \zeta_0/(1+ \omega^2 \tau_0^2)$ \citep{Yang:2023ogo}, which 
reaches a resonant maximum when $\omega=1/\tau_0$ (see, e.g.,
\citet{Harris:2020rus}). Eq.~\eqref{eqn:PpPibulk} is a typical Israel-Stewart-like \citep{Israel:1979wp} equation of motion for the bulk scalar (apart from the source term) \citep{Gavassino:2020kwo}, which is commonly used in the description of the quark-gluon plasma formed in heavy ion collisions \citep{Romatschke:2017ejr} (where $\zeta_0$ and $\tau_0$ are determined by the strong, not the weak interactions). \\

Applications of semi-analytic expressions for the bulk viscosity  \citep{Alford:2019qtm} to the background of a non-viscous neutron star merger calculation have projected that 
large viscous corrections could arise after the merger
\citep{Most:2021zvc}. Here, we go beyond such estimates, as well as
first-order bulk-viscous approximations
\citep{Gavassino:2020kwo,Celora:2022nbp,Camelio:2022ljs}, and instead
compute the full dynamics of the Urca process (in the Fermi Surface
approximation -- see Appendix for an extended discussion) and its influence on dense matter beyond leading order in $\delta \mu$ by directly extracting $\Pi$ from the total pressure \eqref{eqn:PpPi} in our simulations (see also Ref. \citet{Camelio:2022fds,Camelio:2022ljs} for similar simulations  of oscillating neutron stars in spherical symmetry).  We show that
neutron star mergers can reach rapid equilibration and resonant bulk-viscous regimes.

\begin{figure}
    \centering
    \includegraphics[width=0.49\textwidth]{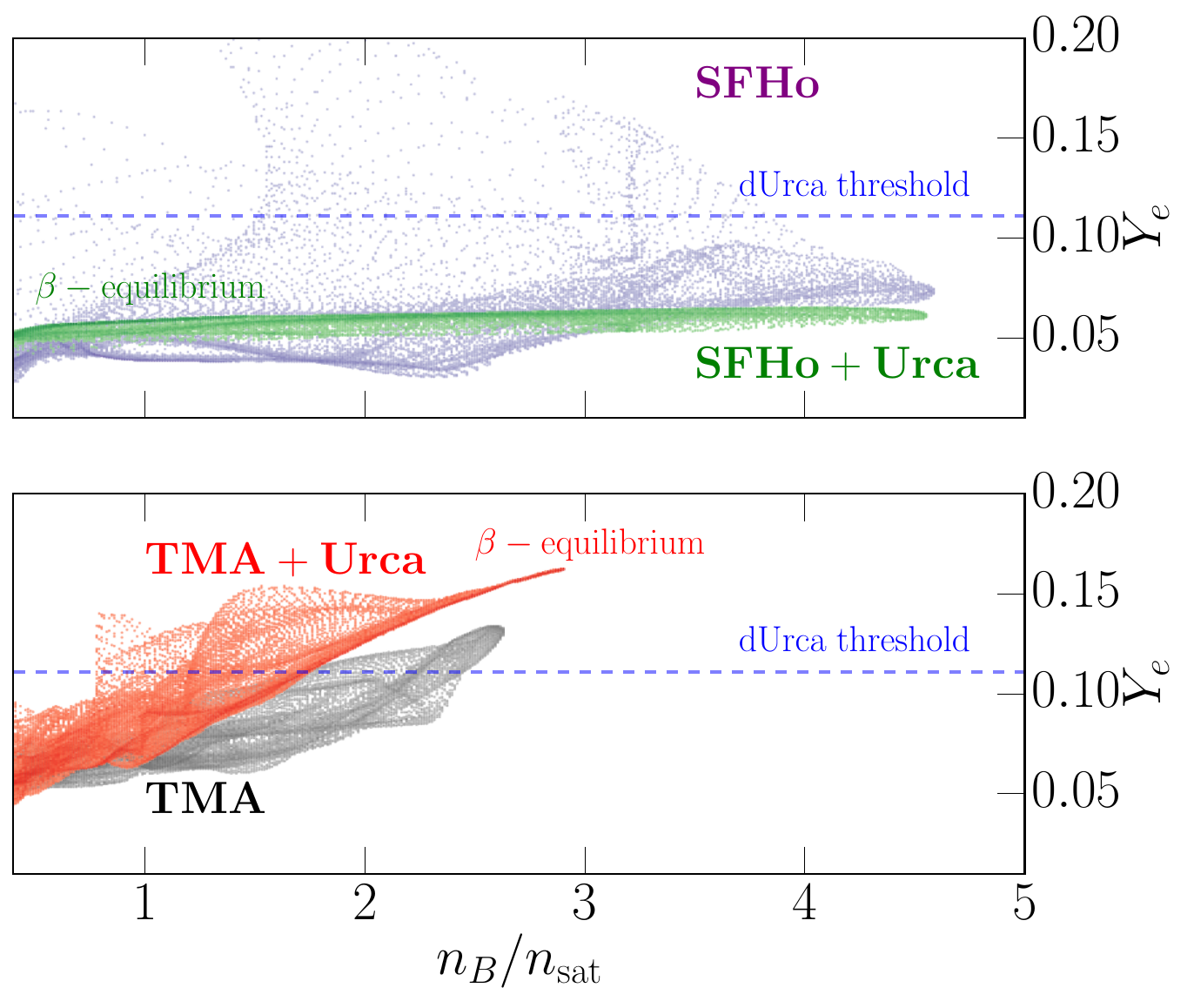}
    \caption{Distribution over density of electron fractions, $Y_e$, present in the system at time $t=5\,\rm ms$ after merger.
    Shown are the results for the simulations using the SFHo (top) and TMA (bottom) equations of state, with and without Urca processes. } 
    \label{fig:Ye_histo}
\end{figure}

\section{Results}

While weak interactions have sufficient time to establish equilibrium for all but the late stages of inspiral \citep{Arras:2018fxj}, the violent dynamics of the merger can drive the electron fraction away from its pre-merger equilibrium value \citep{Perego:2019adq,Hammond:2021vtv}. 
This is displayed in Fig.\ \ref{fig:Ye_histo}, which shows that the distribution of $Y_e$ 
over baryon density 5\,ms after merger
is quite different when Urca reactions, which restore equilibrium on a time scale that depends on the density and temperature, are included.

Fig.\ \ref{fig:Ye_histo} shows a clear correlation between  $n_B$ and $Y_e$ in $\beta-$equilibrium for $n_B> n_{\rm sat}$. In the case of the TMA EoS, this affects the structure of the star since, in the presence of Urca processes, the matter reaches higher densities, $n_B> 2.5\, n_{\rm sat}$. In the case of SFHo, the merger produces some regions with an excess of neutrons and some with an excess of protons; for TMA  there is mostly an excess of neutrons (see Fig.~\ref{fig:Ye_histo}).
If weak interactions are correctly included, this leads to a rapid onset of neutron $\leftrightarrow$ proton conversion in the star, as shown by the difference between equilibration and non-equilibration $Y_e$ profiles in Fig.\ \ref{fig:Ye_histo}. In the case of SFHo, this will be driven mainly by mUrca, as most fluid elements have $Y_e$ values below the dUrca threshold (Fig.\ \ref{fig:Ye_histo}), whereas for TMA it is a combination of dUrca and mUrca. 
We can also anticipate that this change in structure and composition might
lead to a change in the (early) mass ejection (e.g., \citealt{Rosswog:1998hy,Bauswein:2013yna,Sekiguchi:2015dma,Sekiguchi:2016bjd,Radice:2016dwd,Lehner:2016lxy,Bovard:2017mvn,Radice:2018pdn,Nedora:2019jhl}), but leave the details to future work.

 \begin{figure}
    \includegraphics[width=0.4\textwidth]{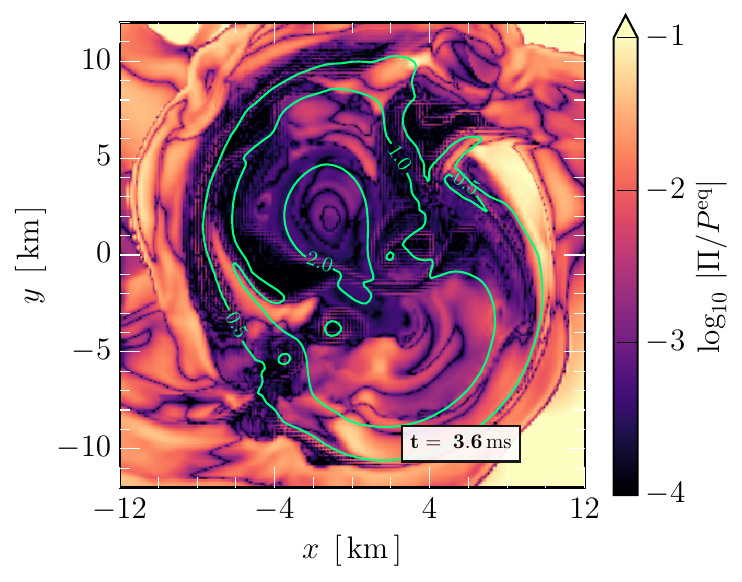}
    \caption{Bulk viscous pressure correction $\Pi$ relative to the equilibrium pressure in the post-merger phase when using the TMA EoS with Urca processes. The green contours denote the rest-mass density of the merger remnant in units of nuclear saturation, at $3.6$ ms relative to the time of merger $t_{\rm mer}$.}
    \label{fig:eff_zeta}
\end{figure}
 
We now move on to the main part of our work, which is to quantify the influence on the global dynamics of beta equilibration via Urca processes.
\begin{figure*}
    \centering
    \includegraphics[width=\textwidth]{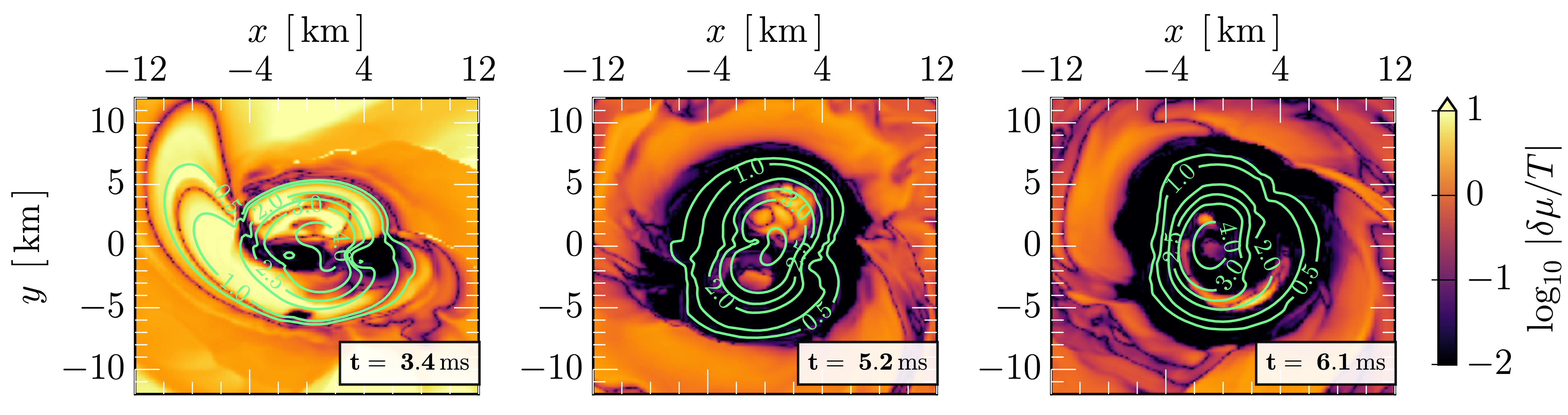}
    \caption{Chemical potential imbalance $\delta \mu$ (normalized by the temperature $T$) as a measure of out-of-$\beta$-equilibrium effects during the merger and post-merger phase of two neutron stars when using the SFHo EoS with Urca processes (the orbital plane is shown). The green contours denote the rest-mass density of the merger remnant in units of nuclear saturation.
    All times, $t$, are stated relative to the time of merger $t_{\rm mer}$.
    }
    \label{fig:eff_press}
\end{figure*}
\begin{figure*}
    \centering
    \includegraphics[width=\textwidth,trim=0 8 0 176 , clip]{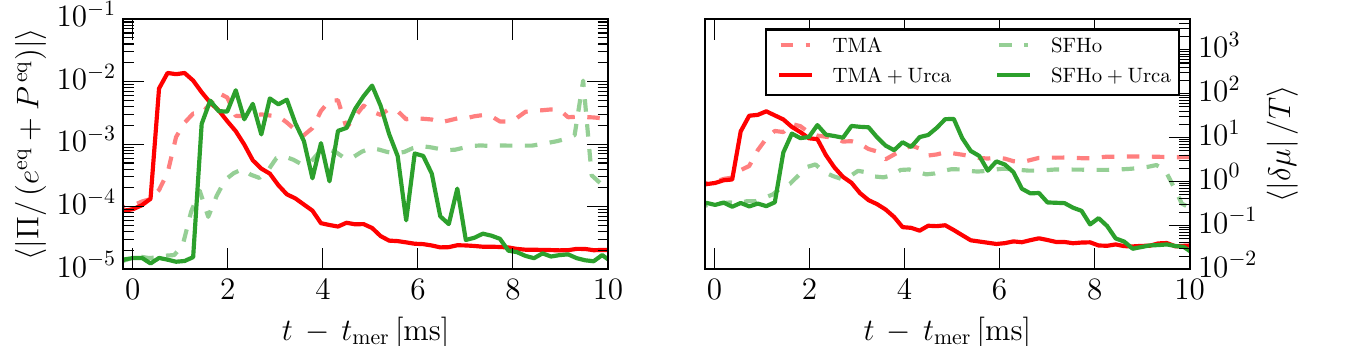}
    \caption{Effective inverse Reynolds number $\chi\equiv\  \Pi/\left(e^{\rm eq}+P^{\, \rm eq}\right)$ and density-averaged chemical potential difference $\delta \mu$ normalized by the temperature $T$ at times $t$ relative to the time of merger $t_{\rm mer}$.
    }
    \label{fig:eff_press_avg}
\end{figure*}
Using our approach, we can provide a first quantification of the out-of-equilibrium dynamics in terms of the relative bulk pressure contribution $\Pi/P^{\rm\, eq}$ as shown in Fig.\ \ref{fig:eff_zeta} for a fixed time. We can see that bulk viscous corrections in terms of $\Pi$ appear in the remnant. We will quantify them in the following and analyze their dynamical impact.

We will also utilize the concept of a dynamic inverse Reynolds number \citep{Denicol:2012cn,Most:2021zvc}, $\chi = \Pi / \left(\ep^{\rm eq}+P^{\rm eq}\right)$, with $\ep^{\rm eq}$ being the total energy density of the fluid in $\beta-$equilibrium. We will, in particular, rely on a density-weighted average value $\left<\chi\right>$.
In Fig.~\ref{fig:eff_press} we show the the departure from $\beta$-equilibrium $\delta\mu$ relative to the temperature: $\delta\mu/T\ll 1$ is the linear response regime.
In the very late inspiral, right before the collision, tidal forces drive the material inside the stars out of $\beta$-equilibrium.
While we only show the beginning of the collision (Fig.\ \ref{fig:eff_press}, left panel), this could potentially set in even earlier \citep{Arras:2018fxj}. Overall, we find that 
$\delta\mu/T \gtrsim 0.1$ in most of the stellar material
(Fig.\ \ref{fig:eff_press}, left panel). The subsequent evolution now strongly depends on whether weak-interaction effects are included.
If Urca processes are not included, 
the matter does not re-equilibrate on the dynamical time scale of the merger and the out-of-equilibrium pressure contribution, $\chi$, remains constant on average for both EoSs (Fig.\ \ref{fig:eff_press_avg}, dashed lines).
If Urca processes are included the system begins to re-equilibrate at a rate that depends on density and temperature. Indeed, we can see (Fig.~\ref{fig:eff_press_avg}, solid lines) that with the inclusion of Urca processes the system relaxes to chemical equilibrium with an average relaxation time comparable to the millisecond timescale of the merger dynamics. If we included neutrino-trapped regions, these would equilibrate much faster \citep{Perego:2019adq}.

During the initial phase of largest out-of-equilibrium dynamics, we see in Fig.~\ref{fig:eff_press_avg} (left panel) that both systems reach $\chi \simeq 0.01$, which is comparable to the effective viscous damping of density oscillations caused by post-merger gravitational wave emission alone  \citep{Most:2021zvc}. In other words, corrections to the macroscopic dynamics can be sizeable in the neutrino free-streaming limit.
Furthermore, in both cases $|\delta \mu|/T \gg 1$, meaning that the system is in a nonlinear far-from-equilibrium regime\footnote{We note that, in the far from equilibrium regime, $\Pi$ is not a linear function of $\delta \mu$ and nonlinear corrections to the equivalence between chemical imbalance and bulk viscosity have to be taken into account \citep{Gavassino:2023xkt,Yang:2023ogo}.} where there is effectively an amplitude-dependent (``suprathermal'') bulk viscosity \citep{Madsen:1992sx,Reisenegger:1994be,Celora:2022nbp,Camelio:2022fds,Camelio:2022ljs}.

However, post-merger oscillations \citep{Stergioulas:2011gd} continue to locally drive the system out of $\beta-$equilibrium. It is especially in this nontrivial far-from-equilibrium regime
 characterized by the approximate plateau in Fig.~\ref{fig:eff_press_avg}, $ u^\mu \nabla_\mu \delta \mu \approx 0$ at $|\delta \mu| /T \gg 1$, where we expect bulk viscous damping to be most pronounced. 
 As a consequence of Eq.\ \eqref{eqn:PpPibulk}, this will also drive an approximate steady state in $\chi$ with fluctuations, which is consistent with Fig.\ \ref{fig:eff_press_avg}.

With this in mind, we can now correlate chemical equilibration and its hydrodynamic feedback in Fig.\ \ref{fig:eff_press_avg}. The equilibration behavior for the two models is qualitatively different:
 For TMA, which is expected to equilibrate faster via dUrca as well as
 mUrca, we see that $\left<\chi\right>$ briefly jumps up after the merger,
 following the no-Urca simulation, but then follows an approximately
 exponential decay with a lifetime of about 1\,ms as Urca processes
 establish $\beta-$equilibrium. For SFHo, which is expected to equilibrate more slowly via mUrca processes, the behavior is more complicated:
 (a) $\left<\chi\right>$ almost immediately reaches a plateau at a value about 10 times higher than in the no-Urca simulations; (b) the plateau persists for about 4\,ms before giving way to exponential decay with lifetime $\sim 1.7$\,ms. 
   
 As a direct consequence, the plateau can only be sustained when matter is driven out of equilibrium by bulk fluid motion at a rate comparable to the flavor relaxation time scale, $\tau$, in this region. From our simulations using the SFHo EoS, we can directly correlate the appearance of these out-of-equilibrium regions with the periodic compressive motion of the stellar cores (see also the regions with $n_B> 3 n_{\rm sat}$ in Fig.\ \ref{fig:eff_press}, center panel).
 
The plateau regions end at $t\approx 5$ ms, which is  when the two cores have fused into a single core, and out-of-equilibrium matter is now found in
lower  density layers ($n_B =2 n_{\rm sat}$, right panel Fig.~\ref{fig:eff_press}). 
  At $t\gtrsim 5$ ms, in the absence of global compressive and expansive motion from core bounces, $\delta\mu$ shows exponential equilibration akin to the TMA case. 

For both EoSs, the flavor relaxation times are shorter than the timescale ($\approx 20-30\, \rm ms$) of post-merger
gravitational wave emission. This implies that for the short-term evolution
of the merger remnant (e.g., \citealt{Fujibayashi:2017xsz,Fujibayashi:2017puw}), the matter can be treated as being in chemical equilibrium.
 As the system approaches chemical equilibrium, $\delta \mu /T \ll 1$, at times $t> 5\, \rm ms$, post-merger oscillations of the stellar remnant only mildly drive the system out of equilibrium, so $\left<\chi\right> \simeq 10^{-5}$ and $\delta \mu /T \simeq 10^{-2}$.

Furthermore, we can qualitatively understand why SFHo matter shows a sustained far-from-equilibrium behavior compared to TMA. The merger drives both systems out of $\beta$-equilibrium, such that there are regions at early times where $|\delta \mu|/T\geq 1$. According to Eq.\ \eqref{eqn:PpPinew}, a non-equilibrium steady-state for $\delta \mu/T$  can only appear in the regions where $\delta\mu +  \mathcal{B}\tau \theta \approx 0$ (assuming nearly constant $T$). This means that the appearance of such a state (and its duration) depends on whether $\mathcal{B}\tau$ can be locally sufficiently large. Given that the expansion rate $\theta$ of both systems is comparable, a reasonable explanation for the approximate plateau in SFHo is that it has a larger $\tau$ because slower modified Urca processes dominate the rates, while TMA reaches the direct Urca threshold, leading to faster equilibration. Consequently, this approximate cancellation effect that drives the non-equilibrium steady state is less prominent in TMA, and Eq.\ \eqref{eqn:PpPinew} predicts the nearly exponential behavior for this EoS seen in Fig.\ \ref{fig:eff_press_avg} in contrast to the approximate plateau seen in SFHo for a few milliseconds.

In the light of the relation between chemical reactions and bulk-viscous processes discussed in \citep{Gavassino:2023xkt}, one may say that SFHo matter in this regime will be in a resummed (i.e., amplitude-dependent) Navier-Stokes-like limit \citep{Yang:2023ogo} allowing us to extract effective bulk viscosities with magnitude $\zeta_{\rm
 eff}\lesssim 10^{28}\, \rm g/(cm\, s)$ directly from the simulation by
 computing $|\zeta_{\rm eff}| \approx| \Pi/ \theta|\approx |\Pi/ \nabla_i u^i|$. We point out that these viscosities can also be calculated precisely in the far-from-equilibrium limit using the same microphysical reactions employed in this work \citep{Yang:2023ogo}. 
  This appearance of dissipation also leads to the production of entropy,
  increasing it on average by $\Delta s\sim 0.1 k_B/\rm baryon$ compared to
  a merger without Urca processes (see also \citet{Most:2022wgo}).

\begin{figure}
    \centering
    \includegraphics[width=0.49\textwidth]{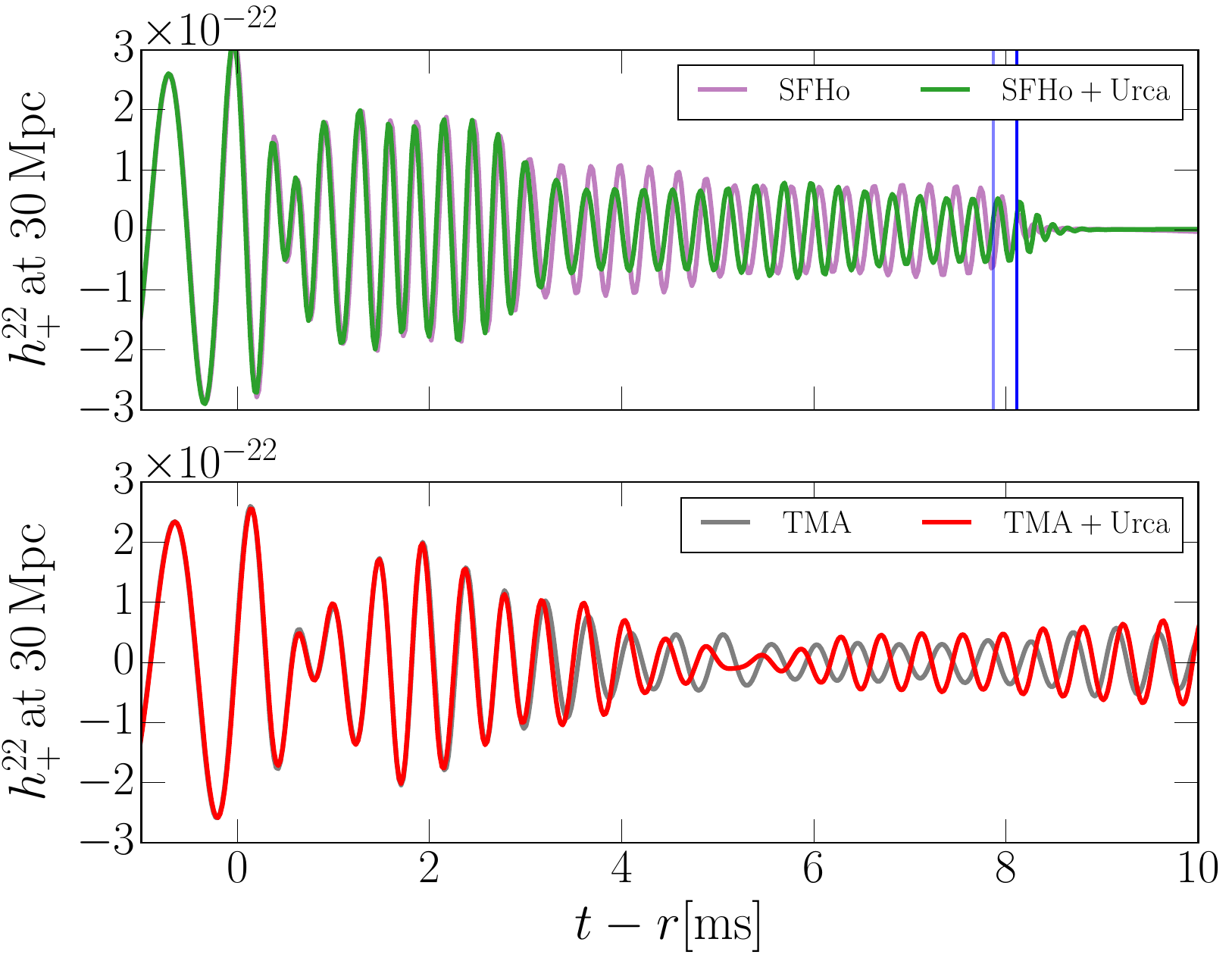}
    \caption{Normalized gravitational wave strain, $h_+^{22}$ for the $l=m=2$ component. Differences due to weak interactions in the post-merger are clearly visible. All times, $t$, are stated relative to the merger time and extraction radius $r$. The vertical lines indicate times of black hole formation.}
    \label{fig:GW}
\end{figure}

Ideally, one would like to identify a signature of dissipative re-equilibration in the gravitational wave signal emitted after the collision. To this end, in Fig.\ \ref{fig:GW} we compare the gravitational emission for all four simulations. For both EoSs, we find good agreement between the gravitational wave strains with and without Urca processes in the late inspiral and the early merger phase. This is consistent with the Urca processes not enforcing $\beta$-equilibrium immediately, as shown in Fig.\ \ref{fig:eff_press_avg}. However, after the characteristic re-equilibration time $\simeq 1-2\, \rm ms$
, a phase difference begins to build up and continues to grow until the end of the simulation $t\simeq 10\, \rm ms$. 

A quantitative estimate of the exact magnitude of how chemical equilibration affects the gravitational wave emission  from our simulation is subject to both effects of numerical resolution and our assumption of neutrino-transparency of the remnant. The latter stems from the fact that flavor equilibration is much slower in the free-streaming regime so there are larger departures from chemical equilibrium. 

In order to provide a complete and faithful quantification of these errors,
we focus on the TMA system, which featured the strongest out-of-equilibrium
effects in the gravitational wave signal. We have performed two additional
simulations at significantly increased numerical resolution $\left( 150
\,\rm m\right)$, one of which imposed the neutrino trapped regime above
temperatures of $T>1\, \rm MeV$, allowing us to quantify both sources of
error (see Appendix). Since the neutrino mean free path depends on the neutrino energy, a clear separation of the neutrino free streaming and neutrino trapped regime by a single transition temperature is not possible. Here, we err on the side of caution and choose a very low cutoff temperature to compute an upper bound on the effect of neutrino trapping.
Based on these simulations, we now comment on potential phase differences between Urca, trapped and frozen composition models.

First, viscous processes are expected to damp out gravitational wave emission.
Using the waveforms shown in the supplemental material, we observe a slight reduction in power for the Urca cases compared to neutrino-trapped and regular evolution cases, providing an indirect consistent indication for a bulk viscous damping of gravitational wave emission. 
Second, dissipation is expected to shift the dominant frequency of post-merger gravitational wave emission.
Our simulations in the neutrino transparent regime (Urca vs.\ frozen composition) feature characteristic shifts of $\Delta f\simeq 40\, \rm Hz$ in both cases. These are roughly comparable to the effects of finite temperature \citep{Raithel:2022orm,Fields:2023bhs}. However, we caution that these are smaller than the resolution error, which we quantify at $70\, \rm Hz$ and thus larger than the shift we measure.
Third, the shift in gravitational wave frequency can also depend on the assumptions on neutrino transparency.
When including neutrino trapping above $T>1\,\rm MeV$ we find an overall
shift of about $50\, \rm Hz$ (again smaller than our numerical error
budget): see also \citet{Hammond:2022uua} for the use of infinitely fast equilibration rates. Additionally, numerical errors can move the shift $\Delta f$ upwards or downwards in these simulations. Ultimately, simulations at higher resolutions and with varying microphysics will be needed to provide a reliable quantification of this effect on the post-merger gravitational wave signal.

\section{Conclusions}
This work presents the first study of weak-interaction-driven out-of-equilibrium dynamics in a binary neutron star merger. We have shown how Urca processes acting on a millisecond timescale restore the departures from $\beta-$equilibrium that arise during the initial stages of the merger. 
We demonstrate the emergence of a microphysical bulk viscosity and
explicitly demonstrate the dissipation induced by Urca processes, and the
establishment of far-from-equilibrium conditions \citep{Gavassino:2023xkt}.
We finally assess the impact on the gravitational wave signal and place
upper bounds on the impact of viscosity and chemical equilibration (see
also  \citep{Hammond:2022uua}). This might change (upwards or downwards) if
significantly finer grids were used (e.g., \citealt{Breschi:2019srl,Kiuchi:2019kzt}).

Further work will be required to fully map out the impact of out-of-equilibrium dynamics on the post-merger evolution. 
Firstly, not all equations of state lead to strong violation of
$\beta-$equilibrium at merger \citep{Most:2021ktk,Hammond:2021vtv}. Those
that do will be subject to uncertainties in the finite-temperature part
\citep{Raithel:2021hye}, critically affecting what fraction of the merger
will be in the neutrino trapped regime
\citep{Perego:2019adq,Alford:2021ogv,Perego:2021mkd}, in which
re-equilibration happens almost instantaneously
\citep{Alford:2019kdw,Alford:2021lpp}. Similarly, trapped neutrino pressure
contributions might be of similar strength, but will affect hot instead of
cold regions \citep{Perego:2019adq}. Furthermore, it remains to be seen what
impact the change in remnant structure has on the mass ejection and
lifetime of the post-merger system \citep{Gill:2019bvq}, see also \citet{Zappa:2022rpd}.
Effects that are not captured by the Fermi Surface approximation, such as finite-temperature blurring of the direct Urca threshold and the resultant modification of the $\beta-$equilibrium condition \citep{Alford:2021ogv,Alford:2018lhf}, should be explored.  Different phases of matter, such as hyperonic and quark matter, have different channels of chemical equilibration giving rise to bulk viscosities with different dependencies on temperature and density \citep{Alford:2020pld,Schmitt:2017efp}.
The presence of a bulk-viscous phase in the merger opens a new window towards connecting dense matter in neutron star mergers to that in heavy-ion collisions \citep{Most:2022wgo}, where out-of-equilibrium viscous corrections are relevant \citep{Romatschke:2017ejr}. 

\section*{Acknowledgments}
ERM thanks F. Foucart, J. Noronha-Hostler, A. Pandya, F. Pretorius, C. Raithel and J. Ripley for insightful discussions related to this work. All authors are grateful to M. Antonelli, S. Bernuzzi, G. Camelio, L. Gavassino, and B. Haskell for very helpful comments on the manuscript, and to L.~Gavassino for pointing out the question of hydrodynamic frame choice associated with Eq. \eqref{eqn:PpPi}.
ERM gratefully acknowledges support from a joint fellowship
at the Princeton Center for Theoretical Science, the Princeton Gravity
Initiative and the Institute for Advanced Study. ERM acknowledges support for compute time allocations on the NSF Frontera supercomputer under grants AST21006. This work used the Extreme Science and Engineering Discovery Environment (XSEDE) \citep{Towns:2014qtb} through Expanse at SDSC and Bridges-2 at PSC through allocations PHY210053 and PHY210074. 
The simulations were also in part performed on computational resources managed and supported by Princeton Research Computing, a consortium of groups including the Princeton Institute for Computational Science and Engineering (PICSciE) and the Office of Information Technology's High Performance Computing Center and Visualization Laboratory at Princeton University. ERM also acknowledges the use of high-performance computing at the Institute for Advanced Study.
ERM acknowledges partial support from the National Science Foundation through PHY-2309210.
MGA, AH, and ZZ are partly supported by the U.S. Department of Energy, Office of Science, Office of Nuclear Physics, under Award No.~\#DE-FG02-05ER41375.
SPH is supported by the U.~S. Department of Energy grant DE-FG02-00ER41132 as well as the National Science Foundation grant No.~PHY-1430152 (JINA Center for the Evolution of the Elements). 
JN is partially supported by the U.S. Department of Energy, Office of Science, Office for
Nuclear Physics under Award No. DE-SC0023861.
ZZ was supported in part by the National Science Foundation (NSF) within the framework of the MUSES collaboration, under grant number OAC-2103680.

\appendix

\section*{Direct Urca rates}
 
In this initial work, we use the Fermi surface approximation to calculate
the Urca rates.  In this assumption, the nuclear matter is treated as
strongly degenerate \cite{Alford:2021ogv,Alford:2018lhf}.  The net rate of
direct Urca is given by (see also \citet{Camelio:2022fds,Camelio:2022ljs})
\begin{equation}
    \Gamma_{\text{dUrca}} \equiv \Gamma_{dU,nd}-\Gamma_{dU,ec} = \frac{G^2(1+3g_A^2)}{240\pi^5}E_{Fn}^*E_{Fp}^*p_{Fe}\theta_{dU}\delta\mu (17\pi^4T^4+10\pi^2\delta\mu^2T^2+\delta\mu^4),\label{eq:durca_net}
\end{equation}
where $G \equiv G_F \cos{\theta_c}$, $G_F=1.166\times10^{-11}\text{MeV}^{-2}$ is the Fermi constant, $\theta_c=13.04^{\circ}$ is the Cabbibo angle, $g_A = 1.26$, and $E^*_{FN} =\sqrt{k^2_{FN}+m_N^{*^2}}$ are the nucleon Fermi energies, and $\delta\mu\equiv \mu_n-\mu_p-\mu_e$ quantifies the departure of the system from chemical equilibrium.  
For simplicity, we fix the effective mass $m_N^{\ast}\approx 0.7\, m_{\rm n}$ relative to the neutron mass $m_n$, although in reality the effective mass drops with increasing density \cite{Alford:2021ogv}.
In the Fermi surface approximation, the direct Urca process operates only above a threshold density, where
\begin{align}
    \theta_{dU}=\begin{cases}
    1 & p_{Fn}<p_{Fp}+p_{Fe}\\
    0 & \text{otherwise}.
    \end{cases}
\end{align}
We neglect finite temperature corrections to the Fermi surface
approximation which would blur this threshold and lead to a nonzero
$\delta\mu^\text{eq}$ \cite{Alford:2018lhf,Alford:2021ogv}.  Above
temperatures of $T\approx 1-10$ MeV we expect neutrinos to be trapped
\cite{Roberts:2016mwj,Alford:2018lhf}. Since the associated Urca timescales
are significantly shorter than the time steps of our simulations
\cite{Alford:2021lpp}, we capture this regime approximately by enforcing
cold $\beta$-equilibrium. Due to the implicit time-stepping, all rates
acting on time scales shorter than the numerical time step, will lead to an
effective instantaneous re-equilibration. Since this happens automatically
for large $T$ we use the same rates also in hot matter.

\section{Modified Urca rates}

The modified Urca rate is also calculated using the Fermi surface approximation.  The individual rates are presented in Sec.~5 of \cite{Alford:2021ogv}.  The net rates of modified Urca, with a neutron or a proton spectator, are
\begin{align}
    \Gamma_{\text{mUrca}, n} &\equiv \Gamma_{mU,nd(n)}-\Gamma_{mU,ec(n)} = \frac{1}{5760\pi^9}G^2g^2_Af^4\frac{(E^*_{Fn})^3E^*_{Fp}}{m^4_{\pi}}\frac{k^4_{Fn}k_{Fp}}{(k^2_{Fn}+m^2_{\pi})^2}\theta_n\\
    &\times \delta\mu (1835\pi^6T^6+945\pi^4\delta\mu^2T^4+105\pi^2\delta\mu^4T^2+3\delta\mu^6).\nonumber
\end{align}
\begin{align}
    \Gamma_{\text{mUrca, p}} &\equiv\Gamma_{mU,nd(p)}-\Gamma_{mU,ec(p)} = \frac{1}{40320\pi^9}G^2g^2_Af^4\frac{E^*_{Fn}(E^*_{Fp})^3}{m^4_{\pi}}\frac{(k_{Fn}-k_{Fp})^4k_{Fn}}{((k_{Fn}-k_{Fp})^2+m^2_{\pi})^2}\theta_p\\
    &\times \delta\mu (1835\pi^6T^6+945\pi^4\delta\mu^2T^4+105\pi^2\delta\mu^4T^2+3\delta\mu^6).\nonumber
\end{align}

Here we have introduced
\begin{align}
    \theta_n=\begin{cases}
    1 & k_{Fn}>k_{Fp}+k_{Fe}\\
    1-\dfrac{3}{8}\dfrac{(k_{Fp}+k_{Fe}-k_{Fn})^2}{k_{Fp}k_{Fe}} & k_{Fn}<k_{Fp}+k_{Fe}\,.
    \end{cases}
\end{align}

and
\begin{align}
    \theta_p=\begin{cases}
    0 & k_{Fn}>3k_{Fp}+k_{Fe}\\
    \dfrac{(3k_{Fp}+k_{Fe}-k_{Fn})^2}{k_{Fn}k_{Fe}} & 3k_{Fp}+k_{Fe}>k_{Fn}>3k_{Fp}-k_{Fe}\\
    \dfrac{4(3k_{Fp}-k_{Fn})}{k_{Fn}} & 3k_{Fp}-k_{Fe}>k_{Fn}>k_{Fp}+k_{Fe}\\
    2+\dfrac{3(2k_{Fp}-k_{Fn})}{k_{Fe}}-\dfrac{3(\textcolor{black}{k_{Fp}}-k_{Fe})^2}{k_{Fn}k_{Fe}} & k_{Fn}<k_{Fp}+k_{Fe}\,.
    \end{cases}
\end{align}

The total modified Urca rate is given by the sum
\begin{equation}
    \Gamma_{\text{mUrca}} \equiv \Gamma_{\text{mUrca}, n} + \Gamma_{\text{mUrca}, p}.
\end{equation}

\section{Total Urca rate}

\begin{figure}
    \centering
    \includegraphics[width=0.49\textwidth]{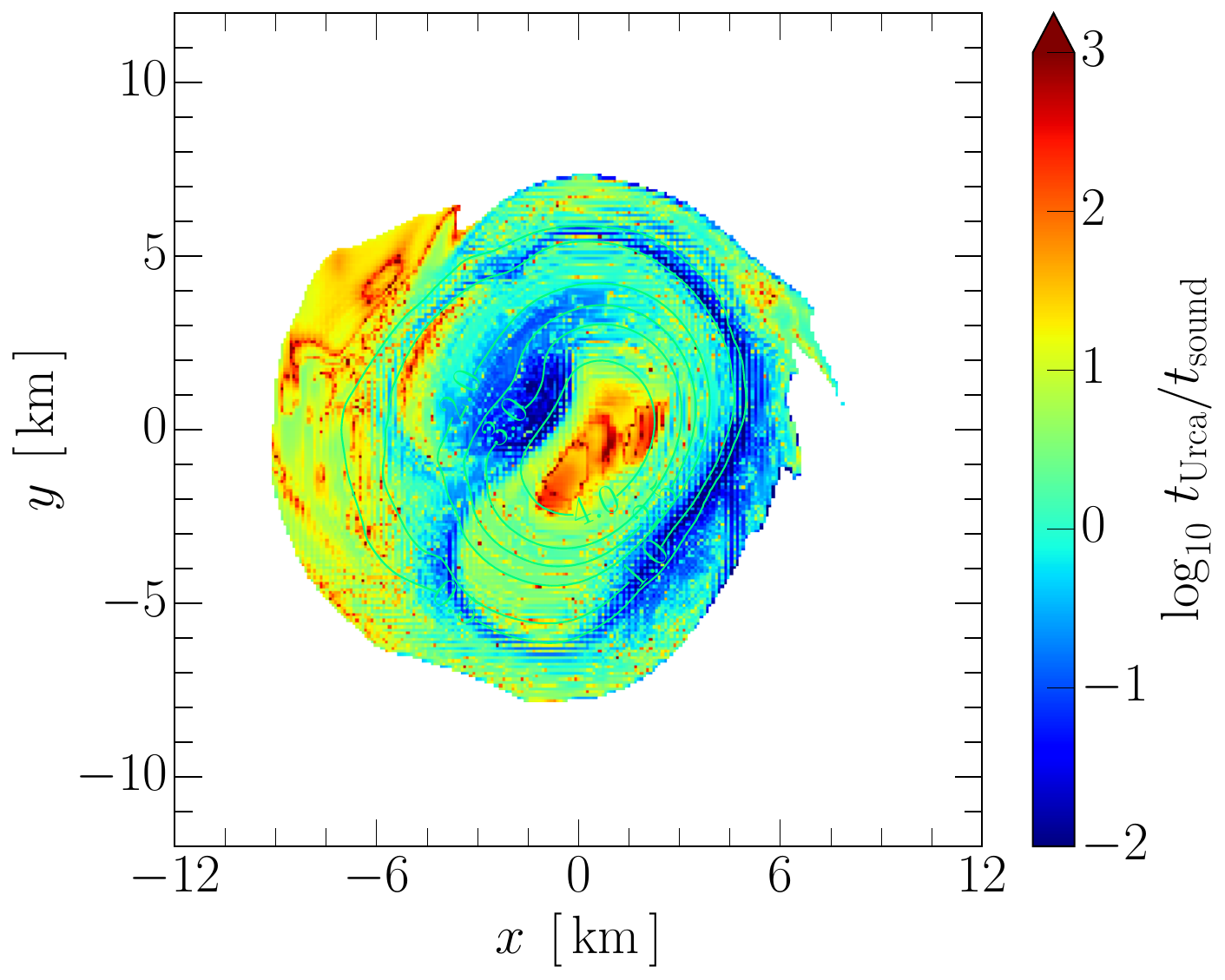}
     \includegraphics[width=0.49\textwidth]{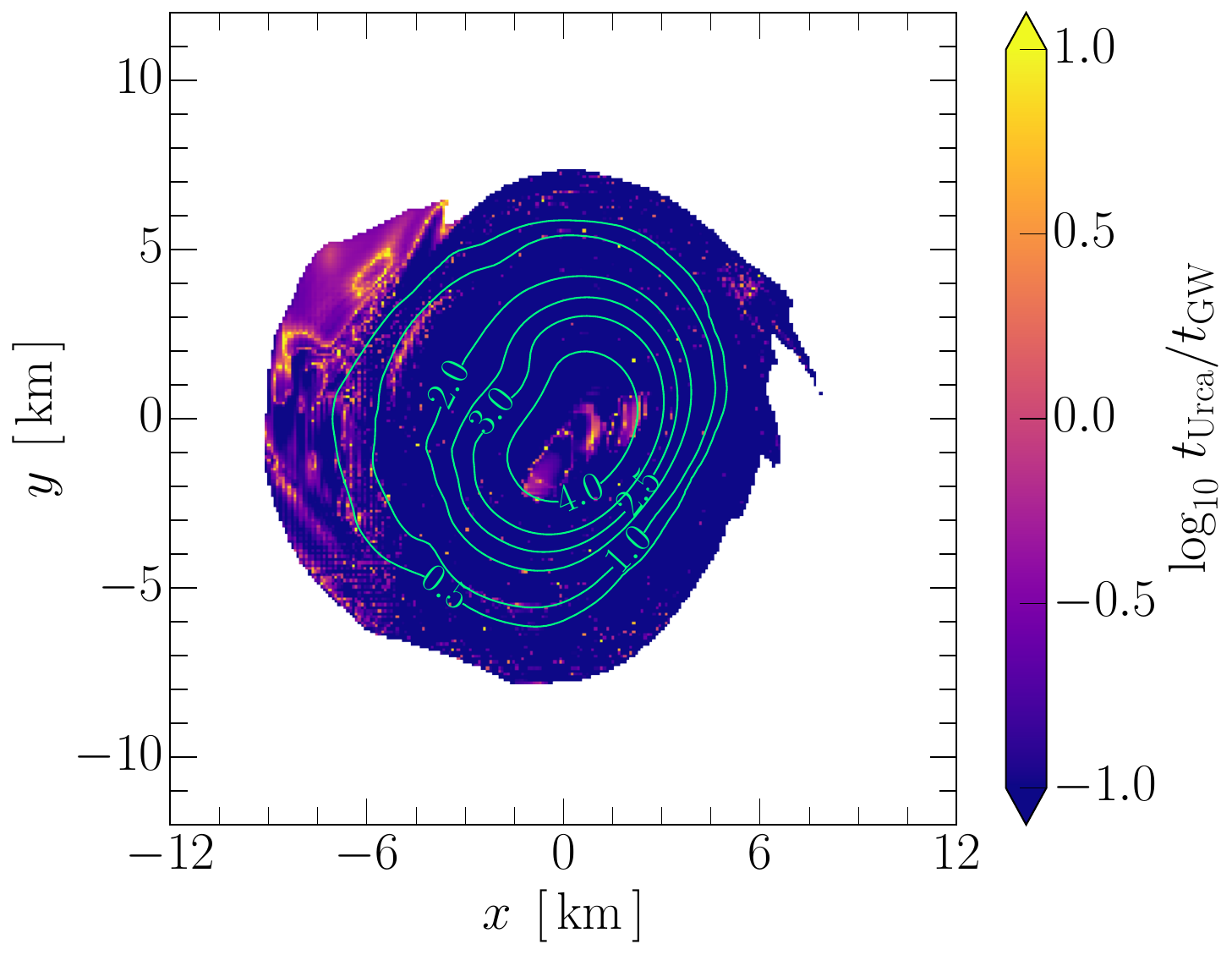}
    \includegraphics[width=0.49\textwidth]{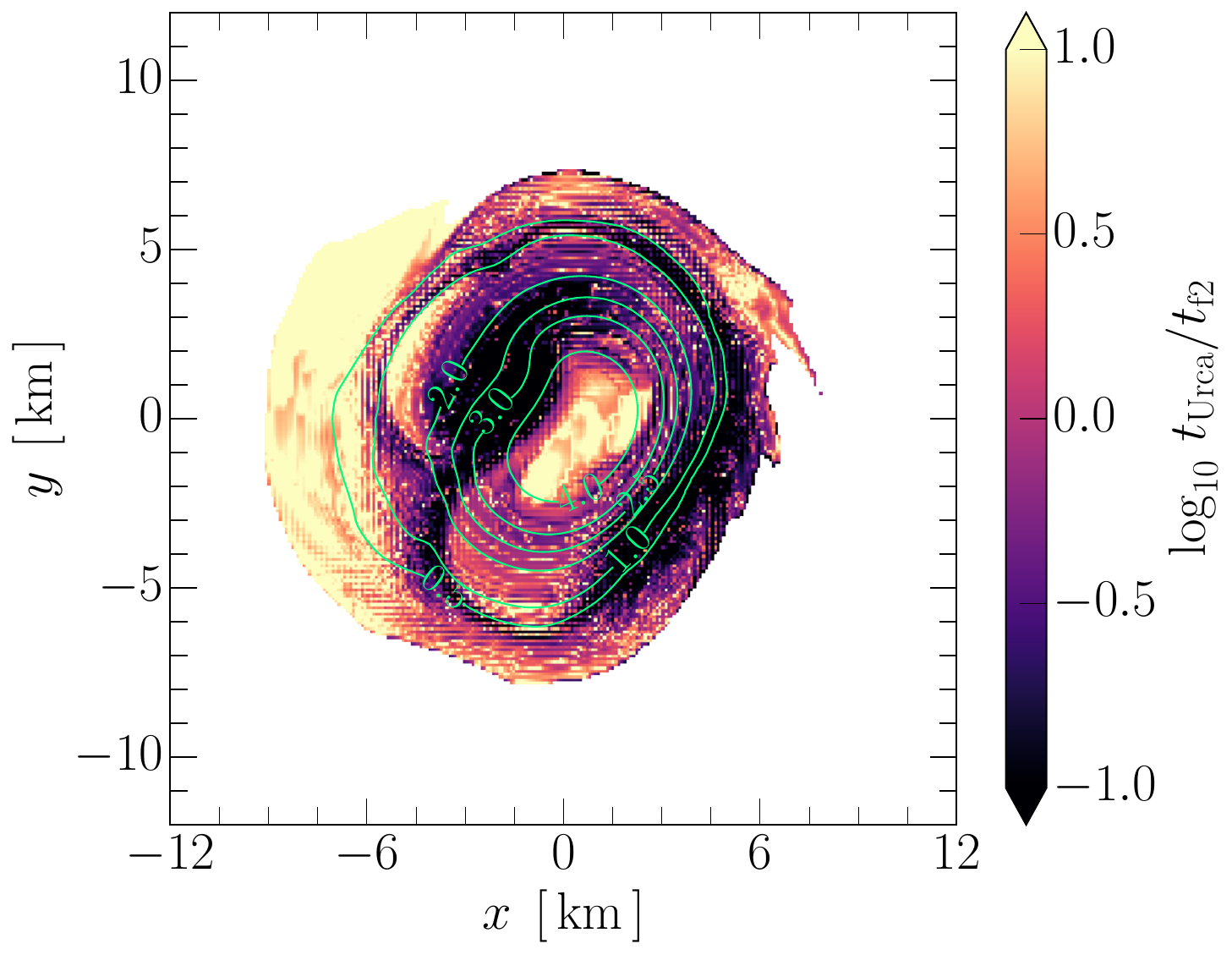}

    \caption{Comparison of the Urca reaction timescale $t_{\rm Urca} = 1/\Gamma$ to the sound crossing time $t_{\rm sound}$, the gravitational wave damping time $t_{\rm GW}$, and the timescale $t_{f2}$ of post-merger gravitational waves, about $6\, \rm ms$ after merger.}
    \label{fig:timescales}
\end{figure}

The total net Urca rate in the Fermi Surface approximation is
\begin{equation}
    \Gamma = \Gamma_\text{dUrca} + \Gamma_\text{mUrca}.
\end{equation}
Since bulk viscosity is typically defined in the linear (also sometimes called ``subthermal'' \cite{Haensel:2002qw,Alford:2010gw}) regime where $\delta\mu\ll T$, it is natural to write this as
\begin{align}
 \Gamma &= \Gamma_\text{bulk}
  + \text{higher order in $\delta\mu/T$} \\
  \text{where}\quad \Gamma_\text{bulk} &\equiv
  \dfrac{d\Gamma}{d\delta\mu}\Bigr|_{\delta\mu=0} \delta\mu. 
 \label{eq:Gamma-bulk}
\end{align}

\section{Comparison of timescales}

In the following, we briefly compare the time scales provided by the Urca reaction rates in the previous section. In order to demonstrate that these rates can indeed affect the dynamics of the post-merger, we compare them to two relevant time scales present in the system.
These effectively govern the propagation of sound waves through the remnant, as well as the characteristic time scale for post-merger gravitational wave emission. In order to bulk viciously damp sound waves, the damping time must be comparable or larger than the sound crossing time, $t_{\rm sound} = d/c_s$. The latter is defined as the time it would take a sound wave to cross the diameter,$d$, of the remnant at its local sound speed $c_s$. We can see in Fig. \ref{fig:timescales}, that this is indeed always the case in cold regions, where the Urca rates act on timescales of about $\simeq t_{\rm sound}$. In hot regions, where the rates equilibrate instantaneously, bulk viscous damping cannot happen. These regions are clearly shown in dark blue, but constitute only a small fraction of the remnant. Similarly, we can show that the Urca rates operate on timescales much faster than the gravitational wave damping timescale $t_{\rm GW} \simeq 30\, \rm ms$ \cite{Most:2021zvc}. This means that bulk viscous damping and re-equilibration can efficiently affect the post-merger gravitational wave dynamics. Finally, we compare the reaction timescale to that of post-merger oscillations in the gravitational wave signal, as quantified by the timescale $t_{\rm f2}\simeq 1/f2$, associated with the dominant peak frequency of post-merger oscillations \cite{Bauswein:2012ya,Takami:2014tva}. We can see that this timescale is similarly comparable across the merger remnant.

\section{Neutrino trapping}

Under conditions probed in the merger neutrinos can become locally
trapped \cite{Perego:2019adq}. In this case, local opacities are so high that the 
neutrinos cannot escape, effectively form a gas, constituting a correction
to the overall pressure, energy density, and lepton number of a fluid
element.\\
In this Section, we will try to provide systematic estimates of the 
effect of neutrino trapping on our results. To this end, we will utilize a
joint post-processing and direct simulation approach, where neutrino
trapping is either estimated or directly included into our simulations
following a similar approach to \cite{Kaplan:2013wra}.

We begin by reviewing the basic description of a neutrino trapping.
In a neutrino trapped fluid element, absorption and scattering will become
as efficient as neutrino emission processes. This implies the formation
of a trapped neutrino gas, which is in chemical equilibrium with the fluid,
\begin{align}
  \mu_{\nu} = \mu_{n} - \mu_p - \mu_e\,,
  \label{eqn:trapped}
\end{align}
for non-vanishing neutrino chemical potential $\mu_\nu$.
While for non-trapped $npe$-matter, the conserved lepton $Y_l$ is given
by the electron fraction $Y_e$, in the case of trapping the total lepton
number is given as the sum of electrons and trapped neutrinos. To this end,
we introduce a trapped neutrino fraction $Y_\nu = Y_l - Y_e =  n_\nu/n_B$, where $n_\nu$
is the trapped neutrino number density and $n_B$ the baryon density.
For a degenerate Fermi gas in thermal equilibrium (i.e., having the same
temperature) such as the trapped neutrino component, we may write
\cite{Kaplan:2013wra},
\begin{align}
  n_\nu = \frac{1}{6\pi^2} T^2 \mu_\nu \left( \frac{\mu^2_{\nu}}{T^2} +  \pi^2 \right)\,. 
\end{align}

Given an initial lepton fraction $Y_l$, which we evolve in our simulations,
we can then compute $Y_\nu$ using the above expression, and obtain 
$\mu_\nu$ from Eq. \eqref{eqn:trapped}. The trapped neutrino component
further exerts a pressure, $P_\nu$. Summed over all three neutrino species,
we find that  \cite{Kaplan:2013wra}
\begin{align}
  P_\nu = \frac{T^4}{24 \pi^2} \left[ \frac{21 \pi^4}{60} +
  \frac{\mu_{\nu}}{2 T^2} \left( \pi^2 + \frac{1}{2}
\frac{\mu_{\nu}^2}{T^2} \right) \right]\,.
\end{align}
We note that due to their massless nature, the energy density $e_\nu = 3
P_\nu$ is trivially related to the pressure.\\

\section{Error budget of the gravitational waveforms}

In the following, we will discuss the error budget of our gravitational wave signals. Since we are interested in comparing signals with different assumptions on chemical equilibration, our error budget is two-fold.
First, all numerical relativity simulations are subject to finite
resolution error. For the code used in this study, this error has been
assessed and quantified in \citet{Most:2019kfe}. In particular, it has been demonstrated that the code is capable of achieving third-order convergence in the post-merger gravitational wave signal, even when using tabulated finite-temperature equations of state. This is particularly important when assessing dynamical aspect of the post-merger phase, such as those considered here. Second, our result are naturally subject to assumptions on neutrino transparencies, with the free-streaming assumption used in this work leading to enhanced chemical equilibration effects compared to a fully neutrino trapped regime.\\

In order to quantify these errors we have performed a new set of simulations for the system using the TMA EoS, which had displayed the largest differences in the gravitational wave signal. For all these simulations, we have adopted a resolution of $150\, \rm m$, which is significantly higher than our standard $256\, \rm m$ resolution used previously.
One of these new simulations applies a neutrino trapping correction (as outlined above) at densities $n> 0.5\,\rm n_{\rm sat}$ and temperatures $T>1\, \rm MeV$, roughly consistent with previous estimates in the literature \cite{Perego:2019adq}. This allows us to assess both errors independently.

\begin{figure}
    \centering
    \includegraphics[width=0.7\textwidth]{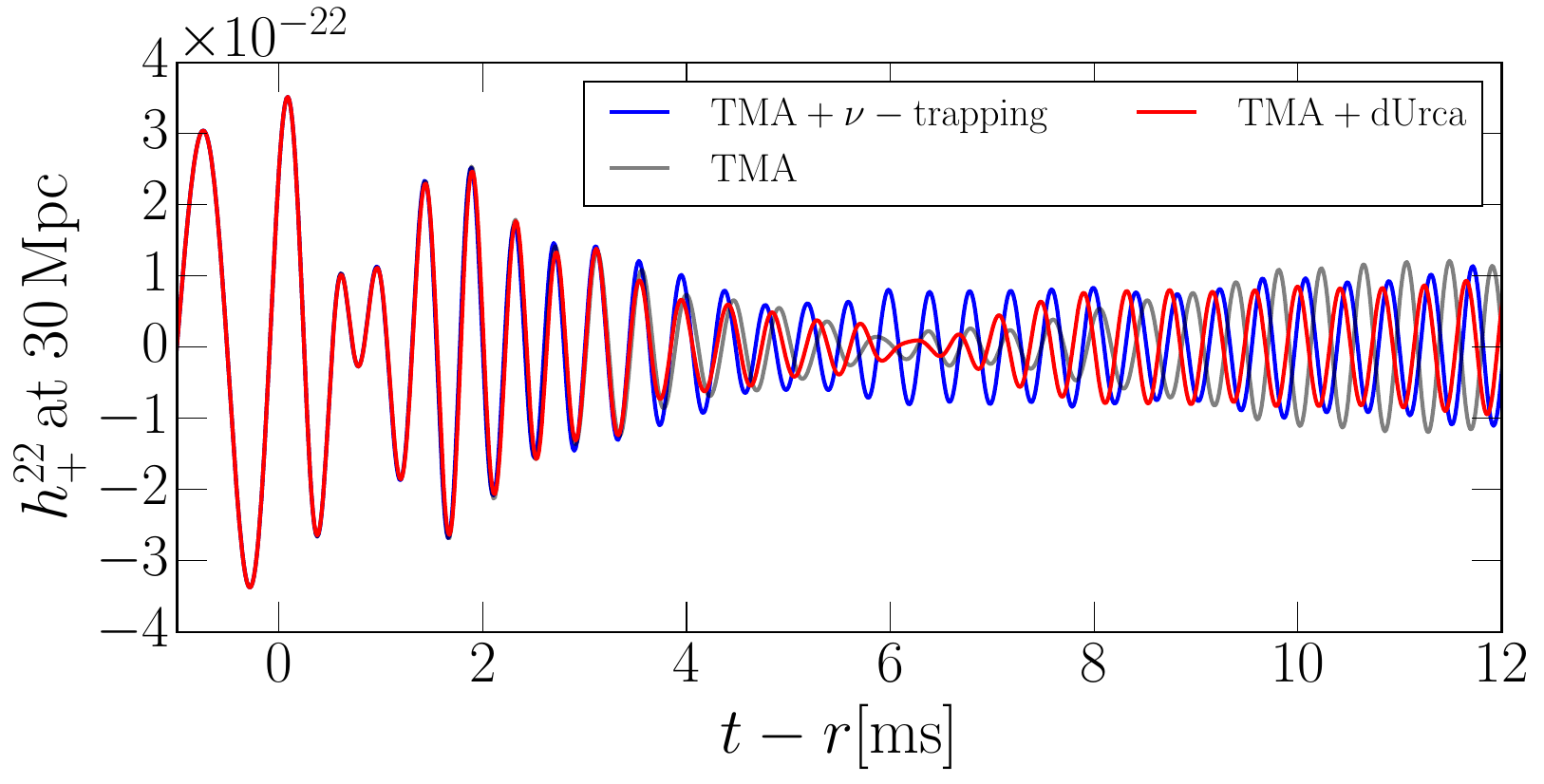}\\
        \includegraphics[width=0.7\textwidth]{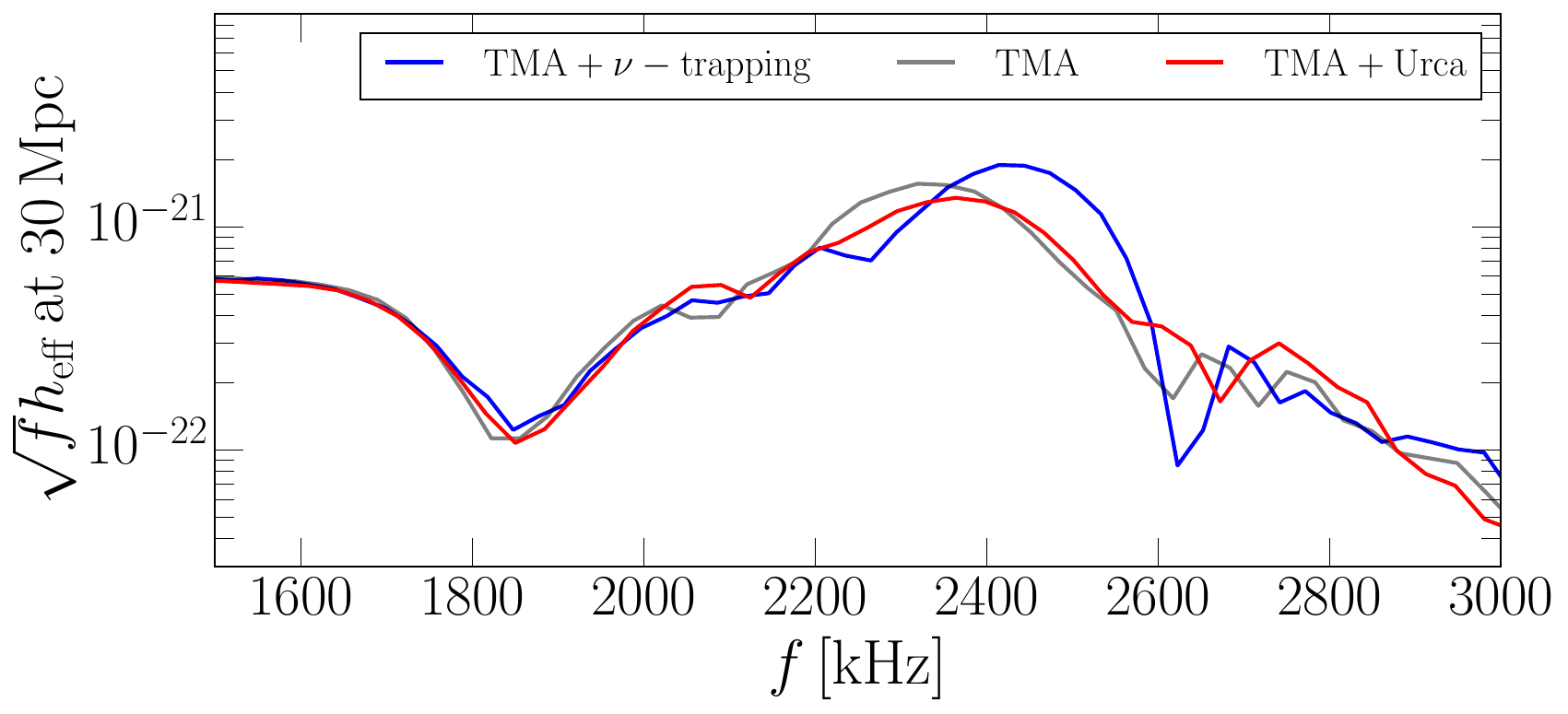}
    \caption{Gravitational wave strain $h_+^{22}$ ($l=m=2$ component) for the merger performed using the TMA equation of state. Different curves refer to various degrees of neutrino trapping. The bottom panel shows the frequency domain waveform around the dominant mode. A bulk viscous suppression in power for the Urca simulation is visible.}
    \label{fig:app_GW}
\end{figure}

The resulting gravitational waveforms are shown in Fig. \ref{fig:app_GW}. We can see a few milliseconds after merger, chemical equilibration leads to a dephasing of these waveforms, initially aligned at merger. For each of these waveforms we compute the power spectrum and identify the dominant peak in post-merger gravitational wave emission \cite{Bauswein:2012ya}. We find differences in those frequencies of $\Delta f \approx (90, 40, 110)\, \rm Hz$, between a) trapping and no equilibration at high resolution, b) free-streaming and no-trapping at high resolution, and c) free-streaming and no equilibration at standard resolutions. For comparison, at standard resolutions, we find $\Delta f \simeq 40\, \rm Hz$ for the SFHo simulations. Furthermore, we also show the power spectrum for the high-resolution TMA waveforms, which feature suppressed power in the $f_2$ mode for the free-streaming Urca case (red curve), which is a clear sign of bulk viscous damping, which can only be active in this case since it requires a dynamical relaxation of the chemical composition, rather than a fixed equilibrium.

\bibliography{inspire,non_inspire}%

\end{document}